\documentclass[twocolumn,showpacs,amsmath,amssymb,prb]{revtex4}

\usepackage{graphicx}
\usepackage{dcolumn}
\usepackage{bm}

\begin{document}

\preprint{APS/123-QED}
\title{Generic Seebeck Effect From Spin Entropy}

\author{$^{1,2,3}$Peijie Sun}
 \email{pjsun@iphy.ac.cn}
\author{$^{1}$K. Ramesh Kumar}
\author{$^{1,2}$Meng Lyu}
\author{$^{1,2}$Zhen Wang}
\author{$^1$Junsen Xiang}
\author{$^4$Wenqing Zhang}
\affiliation{%
$^1$Beijing National Laboratory for Condensed Matter Physics, Institute of Physics, Chinese Academy of Sciences, Beijing 100190, China \\
$^2$University of Chinese Academy of Sciences, Beijing 100049, China \\
$^3$Songshan Lake Materials Laboratory, Dongguan, Guangdong 523808, China \\
$^4$Department of Physics, Southern University of Science and Technology, Shenzhen, Guangdong 518055, China
}%

\date{\today}

\begin{abstract}

How magnetism affects the Seebeck effect is an important issue widely concerned in the thermoelectric community yet remaining elusive. Based on a thermodynamic analysis of spin degrees of freedom on varied $d$-electron based ferro- and anti-ferromagnets, we demonstrate that in itinerant or partially itinerant magnetic compounds there exists a generic spin contribution to the Seebeck effect over an extended temperature range from slightly below to well above the magnetic transition temperature. This contribution is interpreted as resulting from transport spin entropy of (partially) delocalized conducting $d$ electrons with strong thermal spin fluctuations, even semiquantitatively in a single-band case, in addition to the conventional diffusion part arising from their kinetic degrees of freedom. As a highly generic effect, the spin-dependent Seebeck effect might pave a feasible way to efficient ``magnetic thermoelectrics".
\end{abstract}

\pacs{Valid PACS appear here}
\maketitle
The Seebeck effect ($\alpha$) is a key parameter determining the efficiency of useful thermoelectric devices. It measures temperature-difference-induced electric voltage in a conducting solid, being essentially a non-equilibrium thermal transport phenomenon. Nevertheless, in special circumstances this effect can be well depicted thermodynamically by employing thermodynamic state variables. An example is the scaling at low-temperature limit, where electron diffusive transport is suppressed, between $\alpha$ and the electronic specific heat $C_{\rm el}$ $-$ a fundamental thermodynamic property $-$  in a wide range of materials spanning from simple to correlated metals (ref. \cite{behnia04}). Reflecting electron transport kinetics, diffusive characteristics such as the energy-dependent charge mobility can generate a large Seebeck effect (ref. \cite{sun15}) too, as was also revealed by charge-scattering engineered thermoelectricity in nano-scaled materials \cite{martin09}.

Recent years have seen increasing efforts to pursue large values of $\alpha$ in magnetic materials \cite{tsujii19,zhao17,zheng19}. A general knowledge obtained so far is that spin-dependent Seebeck effect (SdSE) might be a material-specific property related to either magnon drag, spin fluctuation or spin-dependent scattering. No generic SdSE has been known. Below, we will demonstrate for a large number of $d$-electron based magnetic conductors with either ferromagnetic (FM), weakly ferromagnetic (WFM), or antiferromagnetic (AFM) transition that a sizable SdSE takes place over a wide temperature range from slightly below to well above the ordering temperature. This additional Seebeck effect, denoted as $\alpha_m$, traces back to the thermodynamics of spin degrees of freedom, i.e., spin entropy, via thermal spin fluctuations of (partially) delozalized magnetic $d$-electrons, cf. the comparison of the Seebeck effect between nonmagnetic [Fig. 1(a)] and magnetic [Fig. 1(b)] conductor. In the latter case, delocalized $d$ electrons are responsible for both collective magnetism and thermoelectric transport. With spin entropy being a principal origin, SdSE is expected to respond to magnetic field scaling to the magnetocaloric effect, a thermodynamic property where field-induced spin-entropy change is feasible to make effective solid cooling, see Figs. 1(c$-$d).

In a conducting solid exposed to temperature gradient d$T$, both chemical ($\mu$) and electrical potential ($e \psi$) change thermodynamically at the two sample ends. In a standard setting of thermoelectric measurements, the d$T$-induced voltage measures the difference of electrochemical potential, $\bar{\mu}$ = $\mu$ + $e\psi$. Approximately, the Seebeck effect arises in two parts \cite{apert16,cai06}
\begin{equation}
- \alpha = \frac{1}{eN}\frac{d\mu}{dT} + \frac{1}{N}\frac{d\psi}{dT}.
\end{equation}
Here $e$ is the free electron charge and $N$ the total carrier number. The first term is purely thermodynamic and known as the Kelvin formula \cite{peter10}. The second includes kinetic information arising from charge relaxation processes. It is sometimes called theoretical or effective Seebeck effect \cite{cai06}, as compared to the experimental one detecting both.

From thermodynamic consideration of an electronic system, $\mu$ is related to partial derivative of the total electronic entropy $S$ with respect to $N$, \cite{peter10}
\begin{equation}
d\mu / dT = -dS/dN.
\end{equation}
Eqs. 1$-$2 show that $\alpha$ probes entropy $S$ per charge carrier so long as the details of electron kinetics can be ignored or is of minor importance.
In magnetic conductors, the relevant spin entropy $S_m$ is of interest and $S_m$ = $Nk_{\rm B}$ln$g$, with $k_{\rm B}$ is the Boltzmann constant and $g$ the total number of spin configurations. Hence, the well-known Heikes formula \cite{chaikin}, $\alpha_m$ = $-(k_B/e)$ $\partial$ln\,$g$/$\partial N$, can naturally be obtained. It is applicable to systems of interacting localized electrons and valid at high enough temperatures where all the spin degrees of freedom are active.

\begin{figure}[t]
\includegraphics[width=1\linewidth]{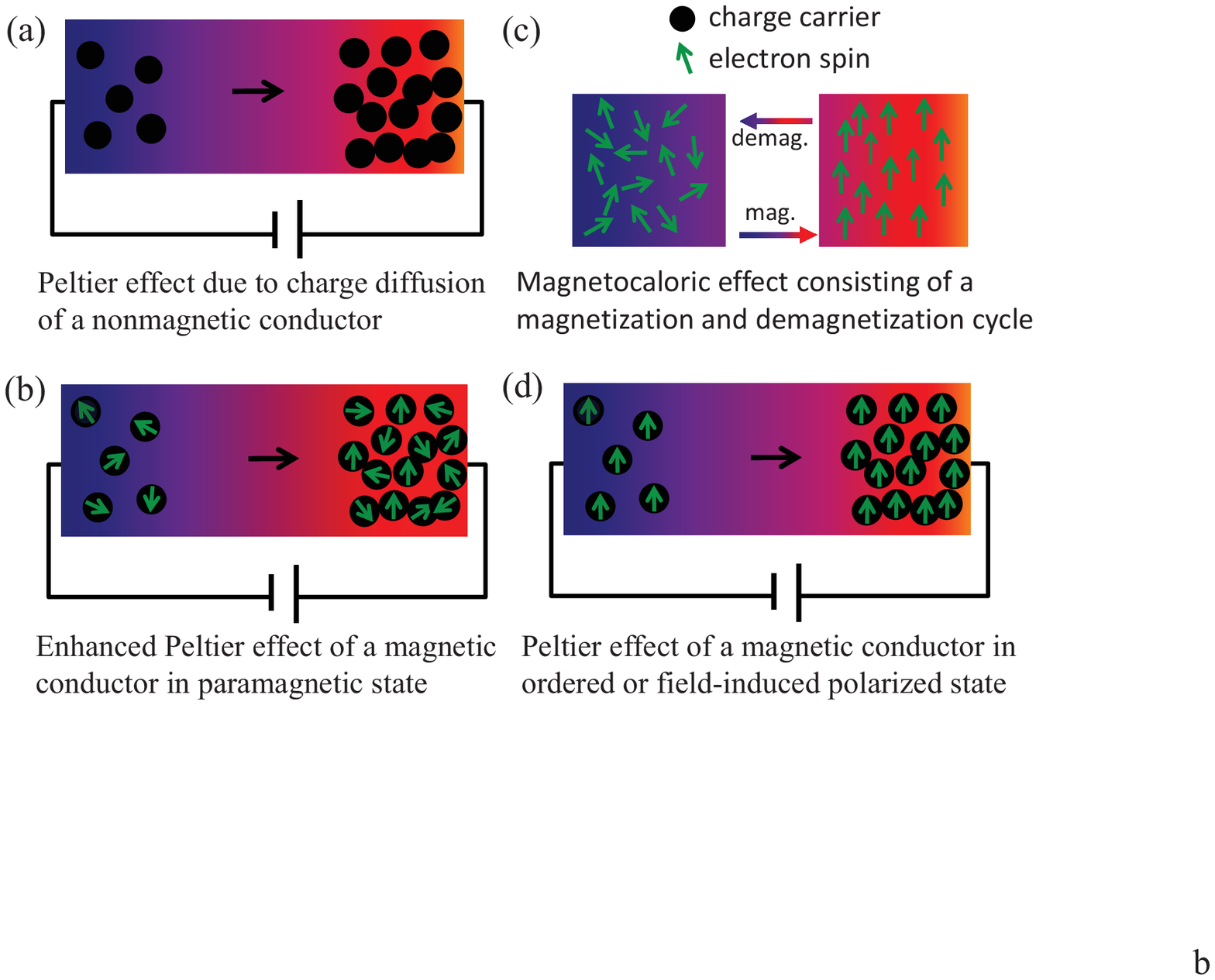}
\caption{Comparison between two relevant solid-cooling effects, the Peltier (the reverse of the Seebeck effect) and the magnetocaloric effect.  (a) The conventional Peltier effect arising from charge diffusion in a nonmagnetic conductor. (b) Enhanced Peltier effect of a magnetic conductor in high-entropy paramagnetic state, involving the thermodynamics of both kinetic and spin degrees of freedom.  (c) A magnetocaloric cycle consisting of magnetization (low entropy) and demagnetization (high entropy) processes of a magnet, where charge carriers are not a necessary ingredient. (d) When sufficiently strong magnetic field is applied or temperature reduced to well below $T_C$/$T_N$, spin polarization occurs and it reduces the spin entropy involved in thermoelectric transport, restoring a situation similar to that of (a).  The difference of the demagnetized (b) and magnetized (d) Peltier effect scales to the magnetocaloric effect, cf. panel (c). Blue and red colors denote heat absorption and release, respectively.
\label{nus.eps}}
\end{figure}

Next we consider the spin entropy of a FM compound on a more general basis within the mean-field approximation \cite{tishin}, whereas AFM state can be represented by two antiferromagnetically coupled FM sublattices.
\begin{equation}
S_m(T, H) = R \left[\ln\frac{\sinh(\frac{2J + 1}{2J} X)}{\sinh(\frac{X}{2J})} - X B_J(X)\right].
\end{equation}
Here $R$ is the molar gas constant, $B_J(X)$ the Brillouin function with $X$ = $g_L$$J$$\mu_{\rm B}$$H_{\rm eff}$/$k_{\rm B}$$T$ and $g_L$ the Lande factor. The effective field $H_{\rm eff}$ reads
\begin{equation}
H_{\rm eff} = H + \frac{3k_{\rm B}T_C B_J(x)}{\mu_{\rm B}g_L(J+1)},
\end{equation}
where the first and second terms at the right-hand side represent external and molecular field, respectively.
Eqs. 3 and 4 are frequently employed to estimate the magnetocaloric effect near magnetic transition \cite{tishin}.

The results of mean-field consideration are summarized in Figs. 2(a$-$d): (a) shows the magnetization per electron $m$ = $g_LJ\mu_B$$B_J(X)$ as a function of $T$ for zero and a finite magnetic field $h$ = 0.2. (b) displays their corresponding $S_m(T)$; it saturates to $R$ln2 = 5.76 J/mol K (the magnetic entropy associated to the ground-state doublet) immediately at $T_C$ for $h$ = 0, whereas the saturation trend is slowed down in finite field. The spin entropy difference $\Delta S_m$ (dashed line) is the magnetocaloric effect and assumes a peak at $T_C$. (c) shows the magnetic specific heat, which usually is the directly measured quantity for estimating $S_m$ and (d) the isothermal field suppression of $S_m$ at selected temperatures. As depicted by Eq. 2, the steplike profile of $S_m(T)$ [Fig. 2(b)] characteristic of magnetic ordering indicates a steplike change of $\alpha(T)$ if the magnetic $d$ electrons are itinerant or partially itinerant, and that $\alpha(H)$ at constant temperature will decrease analogous to $S_m(H)$, see Fig. 2(d).

\begin{figure}[t]
\includegraphics[width=1\linewidth]{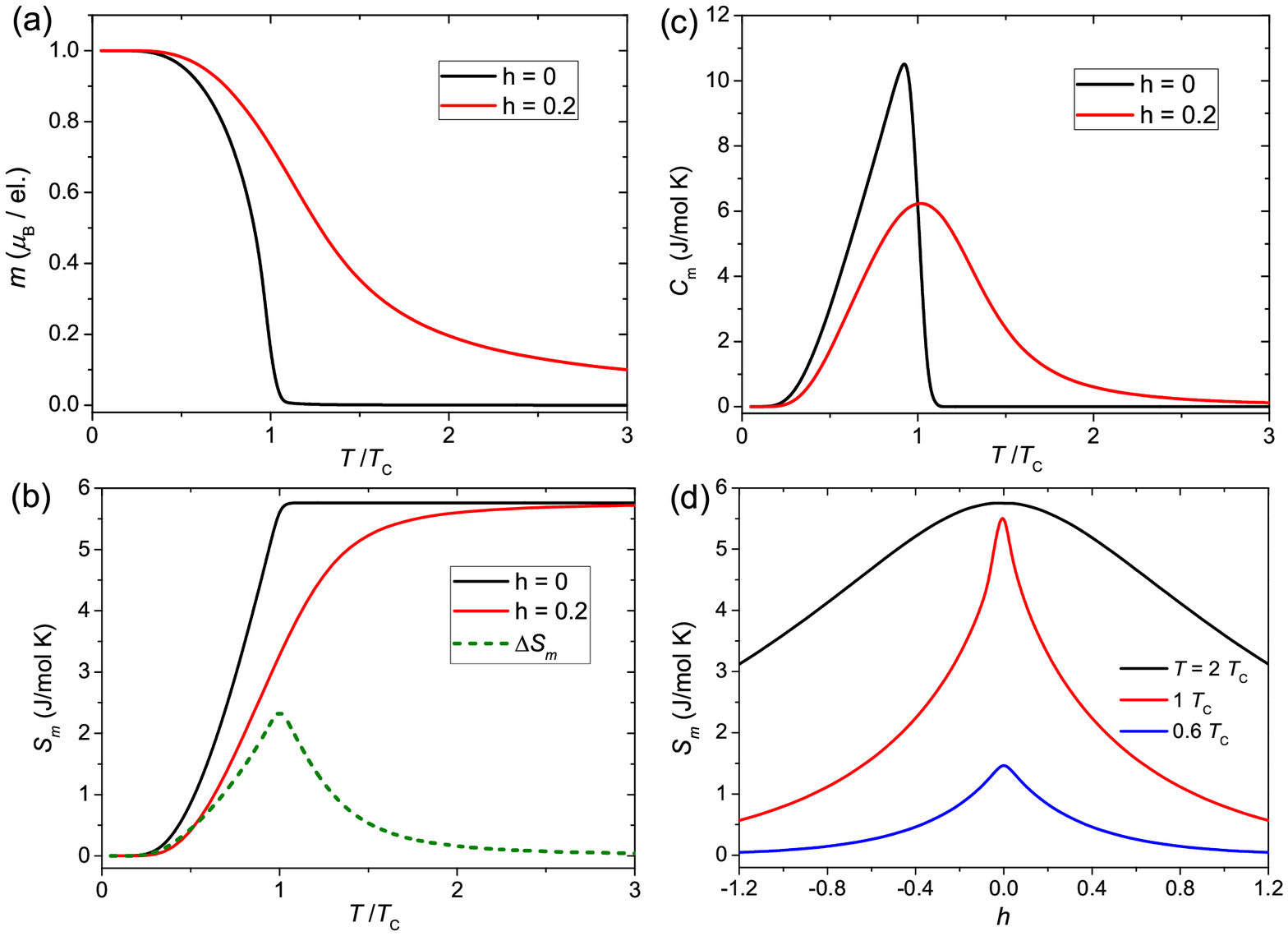}
\caption{Mean-field thermodynamics of a FM spin system with $g_L$ = 2 and $J$ = 1/2. The magnetization $m$ (a), the spin entropy $S_m$ (b), and the magnetic contribution to specific heat $C_m$ (c) are shown as a function of temperature in zero and finite field. Panel (d) display isothermal change of $S_m(h)$ in selected temperatures. In panel (b) the magnetocaloric effect $\Delta S_m$ (dashed line) estimated from the spin-entropy change between zero and finite field is also shown. $h$ = $\mu_B H$/$k_B T_C$ is the reduced magnetic field.
\label{nus.eps}}
\end{figure}

\begin{figure*}[t]
\includegraphics[width=0.85\linewidth]{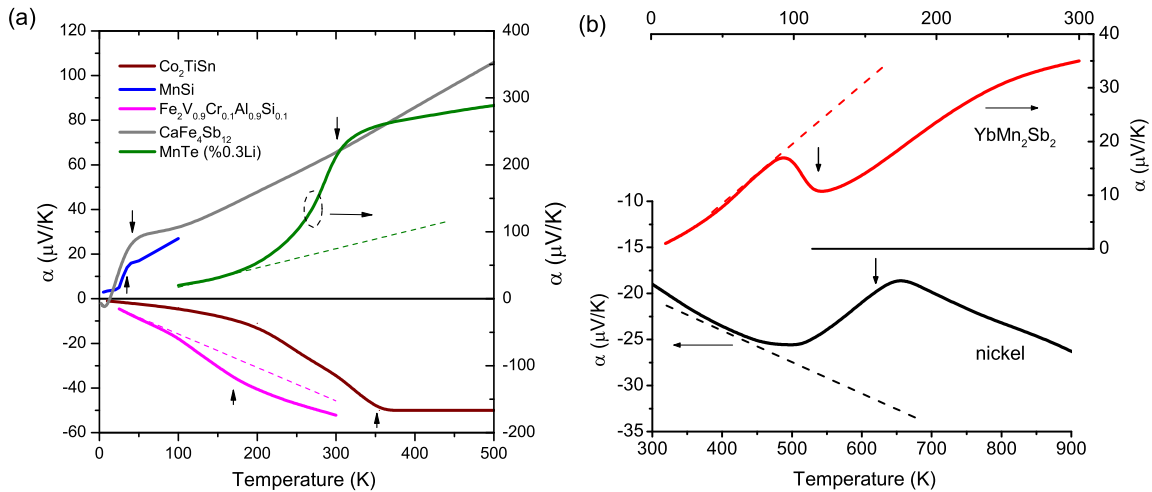}
\caption{Temperature-dependent Seebeck effect of varied $d$-electron FM, WFM and AFM conductors. For all these magnets, a steplike magnetic contribution $\alpha_m(T)$ emerges on top of a sublinear background $\alpha_0(T)$ (dashed line) derived from conventional charge-diffusion effect.  The sublinear background is drawn for only part of the samples for the sake of clarify. Note that, $\alpha_m(T)$ and $\alpha_0(T)$ have same signs in materials shown in panel (a) and opposite signs in panel (b) (see text). The vertical arrows mark the positions of $T_C$ or $T_N$ (for WFM compounds, the position of the maximum in magnetic susceptibility).
\label{nus.eps}}
\end{figure*}

To substantiate the spin-entropy contribution to $\alpha(T)$, in Fig. 3 we compile currently available literature data of various $d$-electron-based magnets, which are either FM, WFM or AFM, see Tab. 1, too. A generic, steplike increase of $\alpha(T)$ at $T$ $\approx$ $T_C$ or $T_N$ is observed for all of them. To avoid complex transport propertied caused by spin-density-wave gap in the AFM cases, samples choice is made for those with a simple resistivity drop below $T_C$/$T_N$, derived from reduced spin scattering. These materials are further classified into two categories: in panel (a) the steplike $\alpha_m(T)$ has the same sign with the background $\alpha_0(T)$, namely, the conventional diffusion contribution; in panel (b), their signs are opposite due to the competing polarity of conduction and magnetic $d$-bands.

\begin{table}[bt]
\caption{\label{tab:table1}
Characterization of the steplike SdSE in some typical $d$-electron based FM, WFM and AFM materials. Among them, Co$_2$TiSn and CaFe$_4$Sb$_{12}$ are representatives of a large group of magnetic Heusler and filled skutterudite compounds, respectively, with similar $\alpha(T)$ profiles, see refs. \cite{balke10,takaba06}.
}
\begin{ruledtabular}
\begin{tabular}{lccccr}
materials & magnetic  & $T_C$/$T_N$ & $\alpha_{\rm step}$ & $M_o$ & refs. \\
 &ordering & (K) & ($\mu$V/K)  & ($\mu_B$) & \\
\colrule
Co$_2$TiSn & FM & 355 & $-$40 &1.97 & \cite{balke10,barth10} \\
MnSi & FM & 29 & 8  & 0.45 & \cite{lamago05,hiro16} \\
Ni  & FM & 627 & 10 &  $\sim$0.6 & \cite{tang71,aba14}\\
CaFe$_4$Sb$_{12}$ & WFM & 50 & $\sim$15 &  0.5 &\cite{takaba06,schne05}\\
Fe$_2$(VCr)(AlSi) & WFM& 160 &7 &  0.4 &\cite{tsujii19} \\
MnTe(0.3\%Li) & AFM & 307 & 150 & 4.55 & \cite{zheng19}\\
YbMn$_2$Sb$_2$ & AFM & 120 & $-$15& 3.6\footnote{In YbMn$_2$Sb$_2$, magnetic moment originates from Mn $d$ bands and divalent Yb ion is nonmagnetic.} & \cite{niki14, moro06}\\
\end{tabular}
\end{ruledtabular}
\end{table}

In spite of the variety of the materials shown in Tab. 1, one sees a rough scaling between the steplike change in $\alpha_m(T)$, $\alpha_{\rm step}$, and the ordered moment $M_{\rm o}$, except for YbMn$_2$Sb$_2$. This hints at a thermodynamic relationship between the two quantities. In-depth investigations into the SdSE of WFM compounds Fe$_2$V$_{1-x}$Cr$_x$Al$_{1-y}$Si$_y$ (ref. \cite{tsujii19}) and CaFe$_4$Sb$_{12}$ (ref. \cite{takaba06}), whose cooperative magnetism has been known to be due to itinerant 3$d$ bands, have shown that applying magnetic field can suppress $\alpha_m$, evidencing the involvement of strong spin fluctuations. The spin-entropy description can readily capture of saturation of $\alpha_m(T)$ above $T_C$/$T_N$ because $S_m(T)$ saturate, or equivalently, the spin degrees of freedom are fully restored. Phenomenologically, paramagnetic spin fluctuations of itinerant $d$ electrons contribute to the Seebeck effect by transferring spin entropy in addition to their kinetic one, see Fig. 1(b).  To verify this proposition, field-tuned $\alpha_m(H)$ will be discussed semiquantitatively in terms of spin entropy $S_m(H)$ [Fig. 2(d)] by focusing on MnSi, see Fig. 4. By contrast, the steplike $\alpha_m(T)$ in Li-doped MnTe has recently been interpreted based on local-moment picture and paramagnon-electron drag \cite{zheng19}. The difficulties arising in the local-moment explanation will be explained below. Here, we note that the electronic structure of this compound is determined by the combined effect of an exchange splitting of Mn-3$d$ states and a strong hybridization with Te-5$p$ states \cite{masek86}. The strong $p-d$ hybridization can partially delocalize the 3$d$ states and cause their contribution to transport \cite{allen77}, appealing for a spin-entropy scenario as well.

Nickel and YbMn$_2$Sb$_2$ shown in Fig. 3(b) are distinctive because their $\alpha_m(T)$ reveals an opposite sign to the sublinear background $\alpha_0(T)$. For nickel, this feature can be explained by considering its ferromagnetism derived from intersite $d$-hole hopping (nickel has an almost full $d$ shell) \cite{okabe94}, which gives rise to a positive spin-entropy contribution on top of the overall negative $\alpha_0(T)$ dominated by the 4$s$ bands at the Fermi level. Aside from the sign problem, the correlation between $\alpha(T)$ and the specific heat $C(T)$ near $T_C$ has long been known for nickel \cite{tang71}, in line with the spin-entropy scenario. Likewise, in YbMn$_2$Sb$_2$, a negative step of $\alpha_m(T)$ appears on a positive $\alpha_0(T)$, due to the competition of magnetic $3d$ electrons of Mn$^{2+}$ and the Sb-5$p$ valence bands. Here, Yb is nonmagnetic.

While not included in Fig. 3, a prominent case where spin entropy has been invoked to interpret SdSE is NaCo$_2$O$_4$ (ref. \cite{tera97}), a paramagnet with Curie-Weiss behavior. Field-induced reduction of $\alpha(H)$ scaling to $H$/$T$, a feature derivable from eq. 3 assuming $T_C$ = 0, has been considered evidence of spin-entropy contribution \cite{wang03}. This argument was rationalized in terms of hopping conduction of local 3$d$ electrons \cite{koshi00}.
Electronic-structure calculation performed later on found this compound locates at an itinerant magnetism instability with large electronic density of states derived from 3$d$ orbitals \cite{singh00}. Within the itinerant picture, the $\alpha(H)$ profile of NaCo$_2$O$_4$ can also be approached by considering magnetic-field tuning to spin polarization  at the Fermi level \cite{singh07}. Here, spin entropy is involved in shaping the electronic structure, as concluded from recent ARPES experiments \cite{chen17}. In this sense, NaCo$_2$O$_4$ appears to be a special case of considerable $d$-electron-derived SdSE with a vanishing $T_C$.

To quantitatively verify the correlation between spin entropy and SdSE, below we scrutinize $\alpha(T, H)$ of MnSi. MnSi is a prototype of itinerant ferromagnet with long-wavelength helical modulation of spin structure. Applying magnetic field reduces the spin entropy near $T_C$, as quantified by the static magnetocaloric effect $\Delta S_m$, see Fig. 2(b) and ref. \cite{arora07}. Consequently, the transport spin entropy detected by $\alpha_m$ is expected to diminish as well, see eq. 2 and Fig. 1(d). Correspondence between the two quantities obtained experimentally is demonstrated in Fig. 4: $\Delta \alpha_m$ ($H$ = 5 T, left axis) reveals a negative peak close to $T_C$ $\approx$ 29 K, see also the inset for the measured $\alpha(T)$ at 0 and 5 T (ref. \cite{hiro16}), from which $\Delta$$\alpha_m$ is obtained as their difference. An apparent scaling between $\Delta \alpha_m$ and $\Delta S_m$ ($H$ = 5 T, right axis, ref. \cite{arora07}) can be observed. Hall-effect measurements reveal a simple one-band normal Hall resistivity with carrier concentration $\sim$ 0.89 $d$-hole/MnSi, along with an anomalous contribution derived from ferromagnetism \cite{neu09}. On this basis, $\Delta S_m$ reported in unit of J/kg K can be readily converted in accordance to the unit of $\alpha$, $\mu$V/K, by normalizing by the carrier density of $d$ holes. The hence obtained spin-entropy-derived Seebeck effect, denoted as $\Delta$$\alpha_{\rm MCE}(T)$ (crosses in Fig. 4), reveals a reasonable agreement with the measured $\Delta \alpha_m(T)$ (squares) within a factor of 2. This yields a compelling evidence that the thermodynamics of delocalized $d$-electron spin degrees of freedom contribute substantially to the SdSE near and  above the magnetic ordering temperature.

\begin{figure}
\includegraphics[width=1\linewidth]{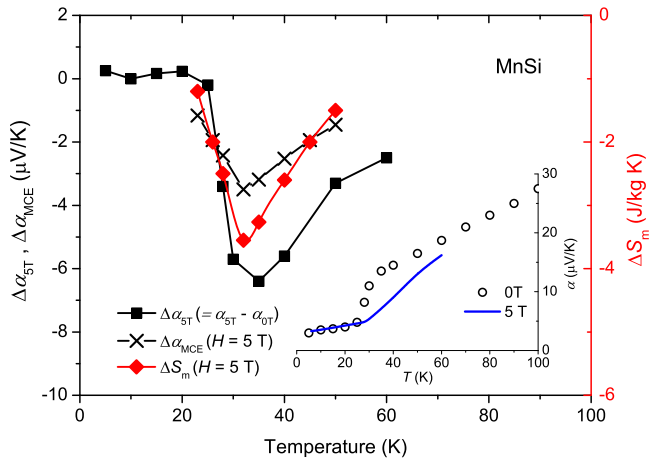}
\caption{Comparison between the measured magneto-Seebeck effect $\Delta$$\alpha_{\rm 5T}$ = $\alpha_{\rm 5T}$ $-$ $\alpha_{\rm 0T}$  (left axis) and the magnetocaloric effect $\Delta$$S_{m}$ for $H$ = 5 T (right axis) for MnSi (ref. \cite{arora07}). By considering 0.89 $d$-hole/MnSi obtained from Hall measurement \cite{neu09}, the values of $\Delta$$S_{m}$ can be converted into $\Delta$$\alpha_{\rm MCE}(T)$ in unit of $\mu$V/K (see text), which agree reasonably with the measured values of $\Delta$$\alpha_{\rm 5T}$. Inset shows $\alpha(T)$ for $H$ = 0 and 5 T (ref. \cite{hiro16}), from which $\Delta$$\alpha_{\rm 5T}$  is obtained.
\label{nus.eps}}
\end{figure}

The SdSE herein identified is generic to magnetic materials that are FM, WFM, AFM or even paramagnetic near a magnetic instability with at least partially itinerant magnetic character. It is a thermodynamic consequence originating from the spin degrees of freedom pertinent to magnetic $d$ electrons. A large number of candidate compounds can be found in literature but are left out of our focus, because their diffusion part $\alpha_0(T)$ is already complex plaguing a reliable analysis of $\alpha_m(T)$. These include, for example,  Fe$_3$GeTe$_2$ (ref. \cite{may16}), the mother compounds of iron-based superconductors \cite{palle16} and various manganese oxides \cite{asa96}. Likewise, a spin-entropy contribution is also expected in transverse thermoelectricity, i.e., the Nernst effect, as recently discussed for a magnetic semimetal Co$_3$Sn$_2$S$_2$ (ref. \cite{gei19}). Unlike MnSi, the transport and static spin entropy may generally deviate from quantitative scaling, because the magnetic $d$ bands may not be the only active bands responsible for transport, or are rather localized as in a magnetic insulator. In the latter case, $\alpha_m$ reduces to zero despite a large $S_m$.

As already mentioned, magnon drag offers an alternative to explain the SdSE, resorting to the kinetics of spin waves interacting with conduction electrons. It in general applies to the ordered phase of local spins \cite{watz16} but has been recently extended to paramagnetic regime far above $T_C$/$T_N$ (ref. \cite{zheng19}). Applying this scenario to $d$-electron based magnets is not straightforward due to their correlated nature with competing localized-itinerant duality. Such magnetic character invalidates the assumption of two physically independent but interacting fluids (conduction electrons and magnons) for the magnon-drag effect. Some even apparent difficulties arise when applying magnon drag to paramagnetic state. One is the sign problem as discussed for Ni: Here, magnon-drag effect appears only well below $T_C$ ($T$ $<$ 500 K) and is known to be negative due to the conduction 4$s$ band, namely, in accordance to the polarity of majority carriers \cite{watz16}. However, $\alpha_m(T)$ near and above $T_C$ is positive [Fig. 3(b)], hinting at a different origin. Also, the large $\alpha_m(T)$ contribution can be observed up to more than 10 times $T_C$/$T_N$ [see, e.g., CaFe$_4$Sb$_{12}$ in Fig. 3(a)], where as far as we know coherent spin fluctuations (magnons) have not been confirmed for any materials. Supposing the steplike $\alpha_m(T)$ is due to paramagnon drag, one would expect a corresponding paramagnon thermal conductivity, too. This has been confirmed in none of the aforementioned materials. Finally, we stress that while spin-entropy scenario does not rely on magnons to drag electrons, thermal spin fluctuations, which offer a standard approach to describe itinerant magnetism \cite{moriya}, are crucial because they are the way by which large spin degrees of freedom are activated.

The physics underlying the $\alpha$$-$$C_{\rm el}$ scaling at 0 K limit in a wide range of correlated compounds \cite{behnia04} can trace back to spin entropy as well. The hybridization between local and conduction bands bring about spin-fluctuation-dressed quasiparticles that are responsible for both transport and thermodynamics, leading to enhanced values of both $\alpha$ and $C_{el}$. This enhancement may, to a certain degree, persist up to even room temperature if the hybridization is strong enough. This is the case of paramagnetic intermediate-valence compound CePd$_3$ (ref. \cite{gam73}), where the optimized figure of merit $zT$ amounts to $\sim$0.3 (ref. \cite{boona12}). By contrast, in itinerant magnets with magnetic ordering as discussed herein, significant spin entropy is drastically accumulated upon approaching the ordering temperature. By choosing magnetic materials with appropriate values of $T_C$/$T_N$, the consequent SdSE can offer more designing flexibility for thermoelectric application in particular temperature range.

In conclusion, based on a thermodynamic analysis we have attributed the excess SdSE near and above $T_C$/$T_N$ in a wide range of $d$-electron-based magnets to spin entropy of the (partially) delocalized $d$ electrons with strong thermal spin fluctuations. The fundamental correlation between the Seebeck effect and spin entropy appears highly instructive for future exploration of useful magnetic thermoelectrics. Furthermore, identification of the physical origin of the spin-dependent Seebeck effect also help to understand the nature of collective magnetism in $d$-electron based systems, which is often not straightforward because of the localized-itinerant duality, and its influence on transport properties.

The authors are grateful to discussion with Y-F. Yang, E. K. Liu, X. S. Wu and G. F. Chen. This work was supported by the National Science Foundation of China (no. 11774404)), and the Chinese Academy of Sciences through the strategic priority research program under the Grant no. XDB33000000.


\begin{thebibliography}{99}
\bibitem{behnia04} K. Behnia, D. Jaccard and J. Flouquet, J. Phys.: Condens. Matter {\bf 16}, 5187 (2004).
\bibitem{sun15} P. Sun, B. Wei, J. Zhang, J. M. Tomczak, A. M. Strydom, M. S{\o}ndergaard, B. B. Iversen, F. Steglich, Nat. Commun. {\bf 6}, 7475 (2015).
\bibitem{martin09} J. Martin, L. Wang, L. Chen, and G. S. Nolas, Phys. Rev. B {\bf 79}, 115311 (2009).
\bibitem{zhao17} W. Zhao $et$ $al.$, Nature {\bf 549}, 247 (2017).
\bibitem{tsujii19} N. Tsujii $et$ $al.$, Sci. Adv. {\bf 5}, eaat5935 (2019).
\bibitem{zheng19} Y. Zheng $et$ $al.,$ Sci. Adv. {\bf 5}, eaat9461 (2019).
\bibitem{cai06} J. Cai and G. D. Mahan, Phys. Rev. B {\bf 74}, 075201 (2006).
\bibitem{apert16} Y. Apertet, H. Ouerdane, C. Goupil, and Ph. Lecoeur, Eur. Phys. J. Plus {\bf 131}, 76 (2016).
\bibitem{peter10} M. R. Peterson and B. S. Shastry, Phys. Rev. B {82}, 195105 (2010).
\bibitem{chaikin} P. M. Chaikin and G. Beni, Phys. Rev. B {\bf 13}, 647 (1976).
\bibitem{tishin} A. M. Tishin, Cryogenics {\bf 30}, 127 (1990).
\bibitem{balke10} B. Balke, S. Ouardi, T. Graf, J. Barth, C. G. F. Blum, G. H. Fecher, A. Shkabko, A. Weidenkaff, and C. Felser, Solid State Commun. {\bf 150}, 529 (2010).
\bibitem{barth10} J. Barth $et$ $al.,$ Phys. Rev. B {\bf 81}, 064404 (2010).
\bibitem{lamago05} D. Lamago, R. Georgii and P. Boni, Physica B {\bf 329-361}, 1171 (2005).
\bibitem{hiro16} Y. Hirokane, Y. Tomioka, Y. Imai, A. Maeda and Y. Onose, Phys. Rev. B {\bf 93}, 014436 (2016).
\bibitem{tang71} S. H. Tang, P.P. Craig, and T.A. Kitchens, Phys. Rev. Lett. {\bf 27}, 593 (1971).
\bibitem{aba14} L. Abadlia, F. Gasser, K. Khalouk, M.Mayoufi, and J. G. Gasser, Rev. Sci. Instru. {\bf 85}, 095121 (2014).
\bibitem{takaba06} T. Takabatake, E. Matsuoka, S. Narazu, K. Hayashi, S. Morimoto, T. Sasakawa, K. Umeo, and M. Sera, Phys. B {\bf 383}, 93 (2006).
\bibitem{schne05} W. Schnelle, A. Leithe-Jasper, M. Schmidt, H. Rosner, H. Borrmann, U. Burkhardt, J. A. Mydosh and Y. Grin, Phys. Rev. B {\bf 72}, 020402(R) (2005).
\bibitem{niki14} V. N. Nikiforov, V.V. Pryadun, A.V. Morozkin, and V.Yu. Irkhin, Phys. Lett. A {\bf 378}, 1425 (2014).
\bibitem{moro06} A. V. Morozkin, O. Isnard, P. Henry, S. Granovsky, R. Nirmala, and P. Manfrinetti, J. Alloys Comp. {\bf 420}, 34 (2006).
\bibitem{masek86} J. Ma\^sek, B. Velick\'y, and V. Jani\^s, J. Phys. C: Solid State Phys. {\bf 20}, 59 (1987).
\bibitem{allen77} J. W. Allen, A. Lucovsky,a nd J. C. Jr Mikkelsen, Solid State Commun. {\bf 24}, 367 (1977).
\bibitem{okabe94} T. Okabe, J. Phys. Soc. Jpn. {\bf 63}, 4155 (1994).
\bibitem{tera97} I. Terasaki, Y. Sasago, and K. Uchinokura, Phys. Rev. B {\bf 56}, R12685 (1997).
\bibitem{wang03} Y.Y. Wang, N.S. Rogado, R.J. Cava, N.P. Ong, Nature {\bf 423}, 425 (2003).
\bibitem{koshi00} W. Koshibae, K. Tsutsui, and S. Maekawa, Phys. Rev. B {\bf 62}, 6869 (2000).
\bibitem{singh00} D. J. Singh, Phys. Rev. B {\bf 61}, 13397 (2000).
\bibitem{singh07} H. J. Xiang and D. J Singh, Phys. Rev. B {\bf 76}, 195111 (2007).
\bibitem{chen17} S. D. Chen, Phys. Rev. B {\bf 96}, 081109(R) (2017).
\bibitem{arora07} P. Arora, M. K. Chattopadhyay, and S. B. Roy, Appl. Phys. Lett. {\bf 91}, 062508 (2007).
\bibitem{neu09} A. Neubauer, C. Pfleiderer, R. Ritz, P. G. Niklowitz, and P. B\"oni, Physica B {\bf 404}, 3163 (2009).
\bibitem{may16} A. F. May, S. Calder, C. Cantoni, H. Cao, and M. A. McGuire, Phys. Rev. B {\bf 93}, 014411 (2016).
\bibitem{palle16} I. Pallecchi, F. Caglieris, and M. Putti, Supercond. Sci. Technol. {\bf 29}, 073002 (2016).
\bibitem{asa96} A. Asamitsu, Y. Moritomo, and Y. Tokura, Phys. Rev. B {\bf 53}, R2952 (1996).
\bibitem{gei19} K. Geishendorf, P. Vir, C. Shekhar, C. Felser, J. I. Facio, J. van den Brink, K. Nielsch, A. Thomas, and S. T. B. Goennenwein, Nano Lett. {\bf 20}, 300 (2020).
\bibitem{watz16} S. J. Watzman, R. A. Duine, Y. Tserkovnyak, S. R. Boona, H. Jin, A. Prakash, Y. Zheng and J. Heremans, Phys. Rev. B {\bf 94}, 144407 (2016).
\bibitem{moriya} T. Moriya, {\it Spin Fluctuations In Itinerant Electron Magnetism} (Springer, Heidelberg, 1985).
\bibitem{gam73} R. J. Gambino, W. D. Grobman, and A. M. Toxen, Appl. Phys. Lett. {\bf 22}, 506 (1973).
\bibitem{boona12} S. R. Boona and D. T. Morelli, Appl. Phys. Lett. {\bf 101}, 101909 (2012).
\end{thebibliography}
\end{document}